\documentclass[aps,prc,twocolumn,showpacs]{revtex4}
\usepackage{graphicx,amsfonts,amssymb,amsbsy}
\usepackage{amsmath}
\usepackage{color,pstricks}

\begin{document}

\title{Surface tension of highly magnetized degenerate quark matter}

\author{G. Lugones$^1$ and A. G. Grunfeld$^{2,3,4}$}
\affiliation{$^1$ Centro de Ci\^{e}ncias Naturais e Humanas, Universidade Federal do ABC, \\ Av. dos Estados 5001, CEP 09210-580, Santo Andr\'{e}, SP, Brazil}
\affiliation{ $^2$ CONICET, Rivadavia 1917, (1033) Buenos Aires, Argentina.} 
\affiliation{$^3$ Departamento de F\'\i sica, Comisi\'on Nacional de Energ\'{\i}a At\'omica, (1429) Buenos Aires, Argentina}
\affiliation{$^4$ Department of Physics, Sultan Qaboos University, P.O.Box: 36 Al-Khode 123 Muscat, Sultanate of Oman}

\begin{abstract}
We study the surface tension of highly magnetized three flavor quark matter within the formalism of multiple reflection expansion (MRE). Quark matter is described as a mixture of free Fermi gases composed by quarks $u$, $d$, $s$ and electrons, in chemical equilibrium under weak interactions. Due to the presence of strong magnetic fields the particles' transverse motion is quantized into Landau levels, and the surface tension has a different value in the parallel and transverse directions with respect to the magnetic field. 
We calculate the transverse and longitudinal surface tension for different values of the magnetic field and for quark matter drops with different sizes, from a few fm to the bulk limit.  For baryon number densities between $2-10$ times the nuclear saturation density, the surface tension falls in the range  $2 - 20$  MeV /fm$^{2}$.
The largest contribution comes from strange quarks which have a surface tension an order of magnitude larger than the one for $u$ or $d$ quarks and more than two orders of magnitude larger than for electrons. 
Our results show that  the total surface tension is quite insensitive to the size of the drop.  We also find that the surface tensions in the transverse and parallel directions are almost unaffected by the magnetic field if $eB$ is below $\sim  5 \times 10^{-3} $ GeV$^2$. Nevertheless, for higher values of $eB$,   there is  a significant increase in the parallel surface tension and a significant decrease in the transverse one with respect to the unmagnetized case. 
\end{abstract}


\maketitle

\section{Introduction}

Systems of strongly interacting matter under the influence of intense magnetic fields are subject of current studies. They have direct application to the physics of neutron stars and the properties of the quark gluon plasma  produced in relativistic heavy ion collisions. In the last case, recent studies indicate that the magnetic field could be as strong $10^{19}$ G \cite{Kharzeev2008}. Some neutron stars, more specifically the so called magnetars, are believed to have large magnetic fields in the core, perhaps as large as  $\sim 10^{19}$ G according to some authors \cite{Duncan1992,Kouveliotou1998,Menezes2009a,Menezes2009b,Avancini2011}.

The cores of massive neutron stars have densities  exceeding a few times the nuclear saturation density. Under such conditions,  a deconfinement transition to quark matter is possible and a hybrid star or a strange quark star can be formed.  The conversion of the star is expected to start with the nucleation of small quark matter drops  \cite{Lugones2010,Bombaci2009,Mintz2010,Hempel2009} which subsequently grow at the expenses of the gravitational energy extracted from the contraction of the object and/or through a strongly exothermic combustion process. Quark matter droplets with a variety of geometrical forms can also arise within the mixed hadron-quark phase that is expected to form inside hybrid stars if global charge neutrality is allowed \cite{Glendenning}. Also, the most external layers of a strange star  may fragment into a charge-separated mixture, involving positively-charged strange droplets (strangelets) immersed in a negatively charged sea of electrons, forming a crystalline solid crust \cite{Jaikumar2006}.

Surface tension is a key ingredient in the understanding of such droplets and the associated phenomenology  \cite{Pinto2012,Wen2010,Palhares2010,Yasutake2014}.  In the case of mixed phases, it is known that there is a critical value for the surface tension of the order of tens of  MeV / fm$^2$ \cite{Voskresensky2003,Tatsumi2003,Maruyama2007,Endo2011} which determines two possible scenarios. If the surface tension is smaller than the critical value, a mixed phase is energetically favored; if it is larger,  the hadron-quark interface in hybrid stars must be a sharp discontinuity. Surface tension is also an important aspect in the process of quark matter nucleation  that leads to the formation of hybrid or strange quark stars because it determines the critical size and the nucleation time of the first quark matter droplets \cite{doCarmo2013,Lugones2011,Buballa2013}.

Even though the surface tension is crucial in the above mentioned phenomena, there is still a broad spectrum of values depending on the models considered for its description. Early estimates pointed to values  below 5 MeV/fm$^2$ \cite{Berger1987}, but larger values within  10 $-$ 50 MeV/fm$^2$ were widely used in the literature, see e.g.  \cite{Heiselberg1993,Iida1998,Bombaci2007,Bombaci2009,Yasutake2014}.
Also, calculations within the linear sigma model with up and down quarks at zero or low temperature and finite quark chemical potential, including vacuum and medium fluctuations predict a surface tension of  $\sim 5 - 15$ MeV/ fm$^2$ \cite{Palhares2010}. However, much larger values have also been reported in the literature. Estimates given in  Ref. \cite{Voskresensky2003} give values in the range 50 $-$ 150 MeV/fm$^2$ and  values around  $\sim$ 300  MeV/fm$^2$ were suggested on the basis of dimensional analysis of the minimal interface between a color-flavor locked phase and nuclear matter \cite{Alford2001}.  More recent works evaluated the surface tension of three-flavor quark matter  within the Nambu-Jona-Lasinio (NJL) model  including finite size effects within the multiple reflection expansion formalism and  the effect of color superconductivity. These calculations also give  large values, above 100  MeV/fm$^2$  \cite{Lugones2011,Lugones2013}.

In this paper we calculate the surface tension of degenerate three-flavor quark matter in chemical equilibrium under weak interactions, assuming that the system is a mixture of free Fermi gases composed by quarks $u$, $d$, $s$ and electrons immersed in a strong magnetic field. We describe finite size effects within the multiple reflection expansion (MRE) framework \cite{Balian1970,Madsen-drop,Kiriyama1,Kiriyama2}. 
The article is organized as follows: in Sect. II we present the quark matter description under the influence of an external magnetic field and include finite size effects. Then, in Sect. III we present our results and finally in Sect. IV give a summary and our conclusions.

\begin{figure*}[tbh]
\includegraphics[angle=0,scale=0.48]{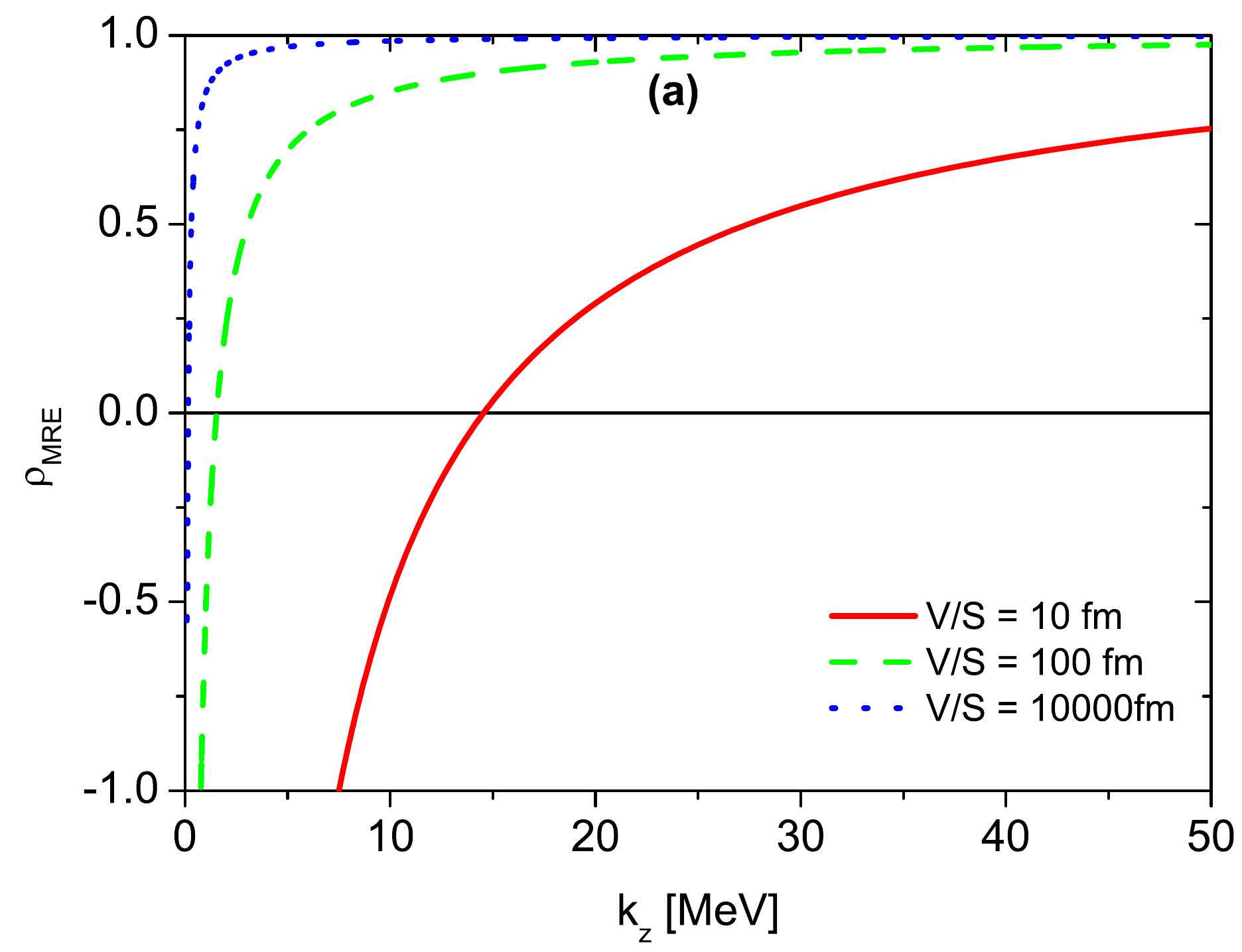} 
\includegraphics[angle=0,scale=0.48]{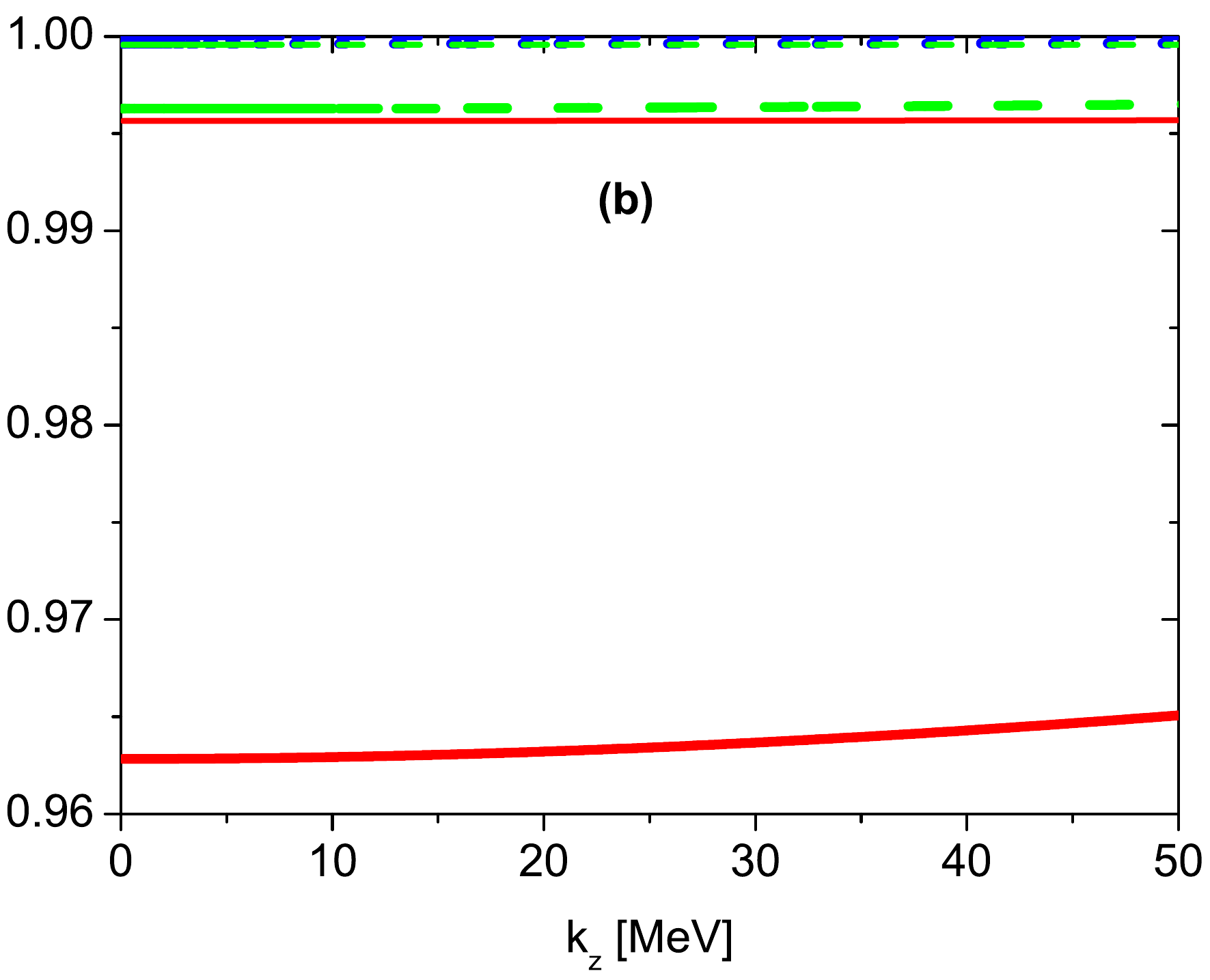}
\caption{The density of states $\rho_{MRE}$ as a function of $k_z$ for the strange quark, using $V/S = 10, 100, 1000$ fm. In panel (a) we show the results for the lowest Landau level $\nu=0$. In such case the curves are the same for any value of the magnetic field. In panel (b) we consider $\nu=1$ with $eB = 0.05$ GeV$^2$ (thick curves) and   $eB = 0.5$ GeV$^2$ (thin curves).}
\label{fig1}
\end{figure*}

\section{Surface effects in a magnetized Fermi gas}

\subsection{Effect of the magnetic field}

Let us consider a Fermi gas immersed in a magnetic field $\textbf{B}$ pointing in the $z$ direction. The transverse motion of particles with electric charge $q_f e$ is quantized into Landau levels (LL) with $k^2_{\perp} = 2 \nu |q_f e B|$, where $\nu \geq 0$ is an integer.  Assuming zero anomalous magnetic moment, the single particle momentum and energy are
\begin{equation}
k =  \sqrt{k_x^2 +  k_y^2 + k_z^2 } =  \sqrt{k_z^2 +  2 \nu |q_f e B| } ,
\label{k}
\end{equation}
\begin{equation}
E = \sqrt{k_z^2 +  m_f^2 + 2 \nu |q_f e B| }.
\end{equation}
Due to Landau quantization, momentum integrals  in the transverse plane must be replaced by sums over the discretized levels. Thus, in the thermodynamic integrals we must use 
\begin{equation}
\frac{1}{(2 \pi)^3} \int \cdots d^3 k \longrightarrow \frac {|q_f e B|}{2 \pi^2} \sum_n \int_{0}^{\infty} \cdots dk_z
\label{Landau2}
\end{equation}
where,  for spin one-half particles, $n$ is related with $\nu$ by
\begin{equation}
\nu = n + \frac{1}{2} - \frac{s}{2}\frac{q}{\left|q\right|},
\end{equation}
being $s= \pm 1 $  the spin projection of the particle along the direction of the magnetic field.

The Fermi momentum for flavor $f$ is 
\begin{equation}
k^{zF}_{f,\nu} = \sqrt{\mu_f^2 - 2 \nu |q_f e B| - m_f^2} . 
\end{equation}
Since $k^{zF}_{f,\nu}$ must be positive,  we obtain a constraint in the sum of Landau levels, 
\begin{equation}
\nu \leq \nu_{max} = \frac{\mu_f^2 - m_f^2}{2 |q_f e B|}.
\end{equation}

According with \cite{deborah} we can replace
\begin{equation}
\sum_n  \rightarrow   \sum_{s = \pm 1}^{} \, \sum_{n=0}^{\nu_{max}} \rightarrow \sum_{\nu=0}^{\nu_{max}} \alpha_{\nu}
\end{equation}
with $\alpha_{\nu} =2$ for all the cases except for ${\nu} = 0$, where $\alpha_{\nu} =1$.
Therefore, we have
\begin{equation}
\frac{1}{(2 \pi)^3} \int \cdots d^3 k \longrightarrow \frac {|q_f e B|}{2 \pi^2}  \sum_{\nu=0}^{\nu_{max}} \alpha_{\nu}
\int_{0}^{\infty} \cdots dk_z
\label{Landau}
\end{equation}
%

\subsection{Finite size effects: inclusion of MRE}

In the present work we consider the formation of finite size
droplets of quark matter. The effect of finite size is included in
the thermodynamic potential adopting the formalism of multiple
reflection expansion (MRE; see Refs. \cite{Madsen-drop,Kiriyama1,Kiriyama2} and references therein).

In the MRE framework, the  modified  density of states of a finite  droplet with an arbitrary shape is given by  \cite{Balian1970}
\begin{equation}
\rho_{MRE}(k,m_f,S, V, \cdots) = 1 + \frac{2 \pi^2}{k} \frac{S}{V} f_S  + \cdots
\label{rho_MRE}
 \end{equation}
where $S$ is the droplet's surface, $V$ its volume, and 
\begin{equation}
f_S(k) = - \frac{1}{8 \pi} \left(1 - \frac{2}{\pi} \arctan \frac{k}{m_f} \right) 
\label{eq:fs}
\end{equation}
is  the surface contribution to the new density of states.  For matter immersed in a magnetic field $\textbf{B}$ pointing in the $z$ direction, the momentum $k$ in Eqs. \eqref{rho_MRE} and \eqref{eq:fs} is given by  $k =  \sqrt{k_z^2 +  2 \nu |q_f e B| }$.  The MRE density of states is shown in Fig. \ref{fig1} for the Landau levels with $\nu=0$ and $\nu=1$, and some values of $V/S$. 

The MRE density of states for massive quarks is reduced  compared with the bulk one, and for a
range of small momenta it may become negative (see Fig. \ref{fig1}). The way of excluding this non physical negative density of states is to introduce an infrared cutoff $\Lambda_{f,\nu}$ in momentum space (see \cite{Kiriyama2} for details). Thus,  we have to perform the following replacement 
\begin{equation}
\frac{1}{(2 \pi)^3} \int \cdots d^3 k \longrightarrow    \frac{1}{(2 \pi)^3}  \int_{\Lambda_{f,\nu}}^\infty \cdots \rho_{MRE} \,   4 \pi k^2 dk.
\label{MRE}
\end{equation}
For magnetized matter, we combine  Eqs. \eqref{Landau} and  \eqref{MRE} and obtain:
\begin{equation}
\frac{1}{(2 \pi)^3} \int \cdots d^3 k \longrightarrow     \frac {|q_f e B|}{2 \pi^2} \sum_{\nu=0}^{\nu_{max}} \alpha_{\nu}  \int_{\Lambda_{f,\nu}}^\infty \cdots \rho_{MRE} dk_z ,
\label{MRE_with_B}
\end{equation}
where $\Lambda_{f,\nu}$ is the cutoff in the momentum along the direction of the magnetic field.

\subsection{Calculation of the infrared cutoff}  
In order to obtain the value of $\Lambda_{f,\nu}$, we have to solve the equation
\begin{equation}
\rho_{MRE}(k_z, m_f,  S, V) = 0
\label{rho_MRE_equal_zero}
\end{equation}
with respect to the momentum $k_z$ and take the larger root as the infrared cut-off. Taking only the first two terms of Eq. (\ref{rho_MRE}) we obtain
\begin{equation}
\frac{k}{m_f}     =  \tan \left( \frac{\pi}{2} - 2 k  \frac{V}{S} \right) .
 \end{equation}
Defining $x = 2Vk/S$ and $\lambda = S / (2Vm)$ the latter equation reads
\begin{equation}
\lambda x = \cot x . 
\label{cutoff_cotangent}
 \end{equation}
Let $x_0$ be the solution of Eq. (\ref{cutoff_cotangent})  for a given value of  $\lambda$. Then, the momentum $k_0$ that verifies Eq. (\ref{rho_MRE_equal_zero}) is given by
\begin{equation}
k_0 = \frac{S}{2V} x_0.
\end{equation}
Using Eq. (\ref{k}) the infrared cutoff on the momentum $k_z$ is obtained from $  \sqrt{k_z^2 +  2 \nu |q_f e B| } = \frac{S}{2V} x_0$; therefore
\begin{equation}
\Lambda_{f,\nu}  =  \sqrt{ \frac{S^2}{4V^2} x^2_0  -   2 \nu |q_f e B| } .
\end{equation}
In Table \ref{cutoff} we show the values of $x_0$ for different quark masses and different values of $V/S$.

\begin{table}[tb]
\centering
\begin{tabular}{c c c | c}
\hline \hline
particles & $m$ [MeV]          &          $V/S$ [fm]              &   $x_0$               \\
\hline  
electrons &0.511     &   10                 &  0.225631  \\   
  &0.511     &    50                &   0.487927  \\   
  &0.511     &    100                &   0.663088  \\   
  &0.511     &    $\infty$                &   $ \pi / 2$ \\   
 \hline 
quarks u, d &5     &   10                 &  0.657008  \\ 
&5     &   50                 &  1.14604   \\ 
&5      &   100                &  1.31661  \\ 
&5      &   $\infty$                 &   $\pi/2$     \\ 
\hline
quarks s &150     &   10                 &  1.47413  \\ 
&150   &   50                 &  1.5504    \\ 
&150  &   100                &  1.56053 \\ 
&150   &   $\infty$                 &  $\pi/2$      \\ 
\hline \hline 
\end{tabular}
\caption{Solution of Eq. (\ref{cutoff_cotangent}) for different particle masses and different values of $V/S$.} \label{cutoff}
\end{table}

\subsection{Parallel surface tension}

Following Eq. (14) of Ref. \cite{Strick}, the pressure of degenerate particles parallel to the magnetic field lines is 
\begin{equation}
{{P}_f^{\parallel} } =    \frac {|q_f e B|}{2 \pi^2} 
\sum_{\nu=0}^{\nu_{max}} \alpha_{\nu} \int_{0}^{k^{zF}_{f,\nu}}  \frac {k_z^2 dk_z}{\sqrt{k^2 + m_f^2}} ,
\end{equation}
with $k =   \sqrt{k_z^2 + 2 \nu |q_f e B| }$.
Including finite size effects in the expression above, we can write the parallel thermodynamic potential of a magnetized quark matter drop within the MRE formalism as:  
\begin{eqnarray}
- {\Omega_f^{\parallel} }& = &    V  \frac {|q_f e B|}{2 \pi^2} 
\sum_{\nu=0}^{\nu_{max}} \alpha_{\nu} \int_{\Lambda_{f,\nu}}^{k^{zF}_{f,\nu}}  \frac {\rho_{MRE}  k_z^2 dk_z}{\sqrt{k^2 + m_f^2}}  \nonumber  \\ 
 & = & \frac { |q_f e B|}{2 \pi^2} 
\sum_{\nu=0}^{\nu_{max}} \alpha_{\nu} \int_{\Lambda_{f,\nu}}^{k^{zF}_{f,\nu}}  \frac {k_z^2 dk_z}{\sqrt{k^2 + m_f^2}}  \nonumber \\
  & &  \times  \left(V + \frac{2 \pi^2 f_S(k)}{k} \times S \right)  .
\end{eqnarray}
The latter expression can be written in the form
\begin{eqnarray}
\Omega_f^{\parallel}  = -  \Pi_f^{\parallel}  V +  \alpha_f^{\parallel}   S 
\end{eqnarray}
where $\Pi_f^{\parallel}$ is the parallel pressure within the MRE formalism and $ \alpha_f^{\parallel}$ is the parallel surface tension, i.e.,
\begin{eqnarray}
\Pi_f^{\parallel} &=&  \frac{|q_f e B|}{2 \pi^2} \sum_{\nu=0}^{\nu_{max}} \alpha_{\nu}  \int_{\Lambda_{f,\nu}}^{k^{zF}_{f,\nu}} \frac{k_z^2 \, dk_z}{\sqrt{k^2 + m_f^2}} ,  \\
\alpha_f^{\parallel} &=& - |q_f e  B| \sum_{\nu=0}^{\nu_{max}} \alpha_{\nu} 
\int_{\Lambda_{f,\nu}}^{k^{zF}_{f,\nu}}   \frac {f_S(k) k_z^2 \, dk_z}{k \sqrt{k^2 + m_f^2}}  .
\end{eqnarray}
In these expressions  $f_S(k)$ is given by Eq. \eqref{eq:fs} and the momentum $k$ by Eq. \eqref{k}.  
To obtain the total parallel surface tension, we have to add the contribution of all particle species, e.g. quarks $u$,  $d$, $s$ and electrons.


\begin{figure*}[tbh]
\includegraphics[angle=0,scale=0.48]{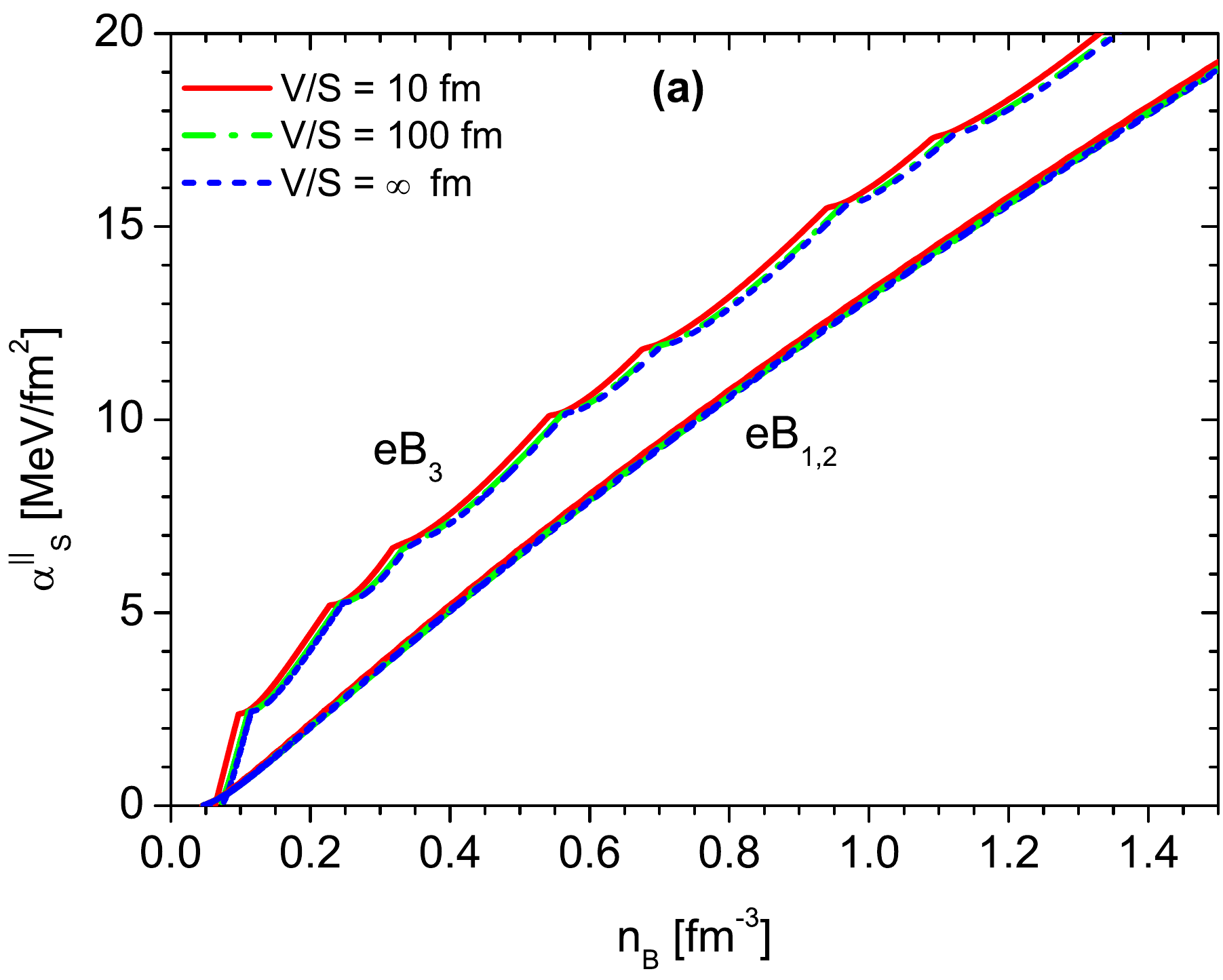}
\includegraphics[angle=0,scale=0.48]{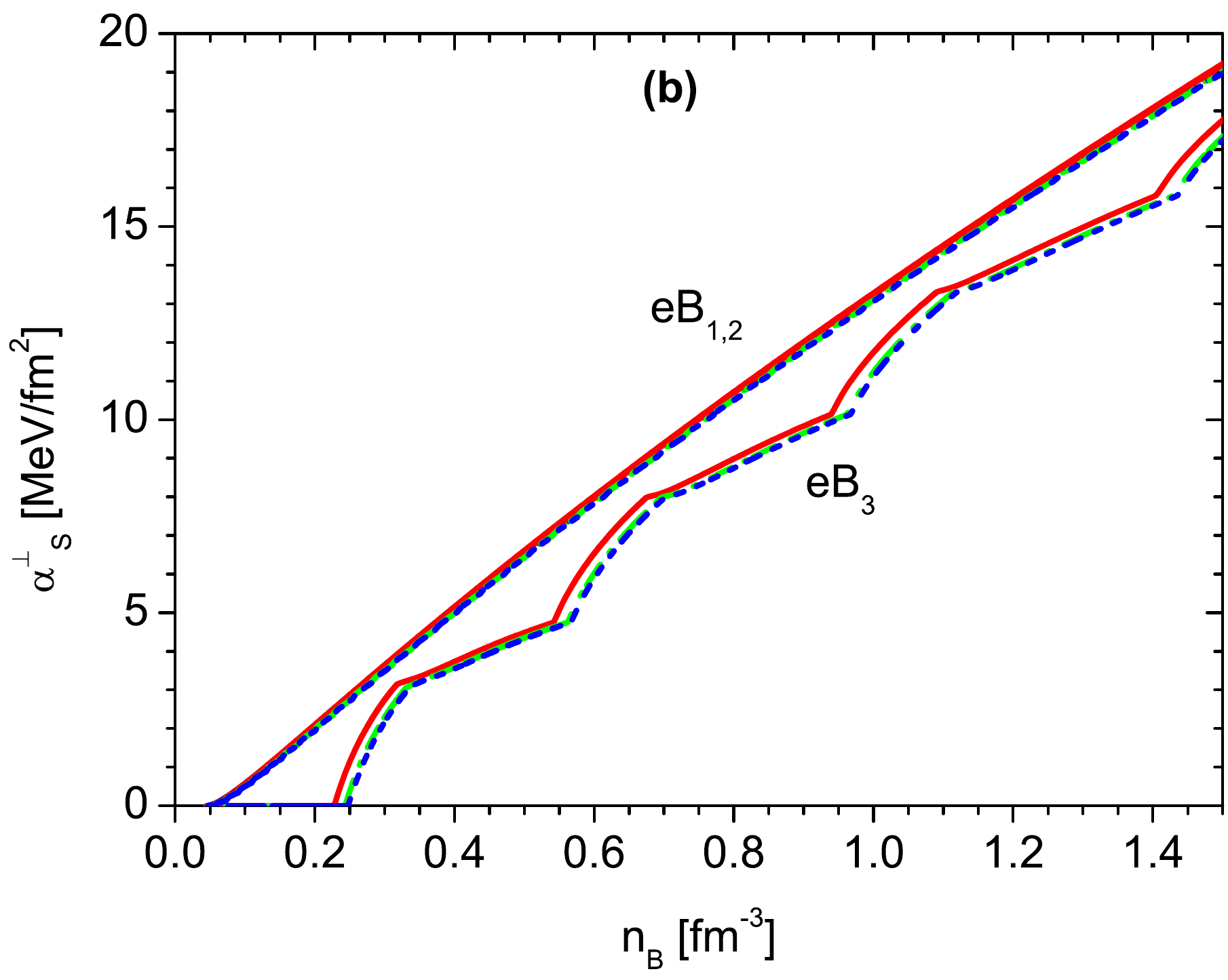}
\caption{Surface tension for quarks $s$  in the parallel direction (panel (a)) and in the transverse direction  (panel (b)).  The results are shown for drops with $V/S$ =  10 fm, 100 fm and  for the bulk limit with $V/S = \infty$. The magnetic field intensities are $eB_1 = 5 \times 10^{-4} $ GeV$^2$, $eB_2 = 5 \times 10^{-3} $ GeV$^2$ and $eB_3 = 5 \times 10^{-2} $ GeV$^2$. }
\label{fig2}
\end{figure*}

\begin{figure*}[tbh]
\includegraphics[angle=0,scale=0.54]{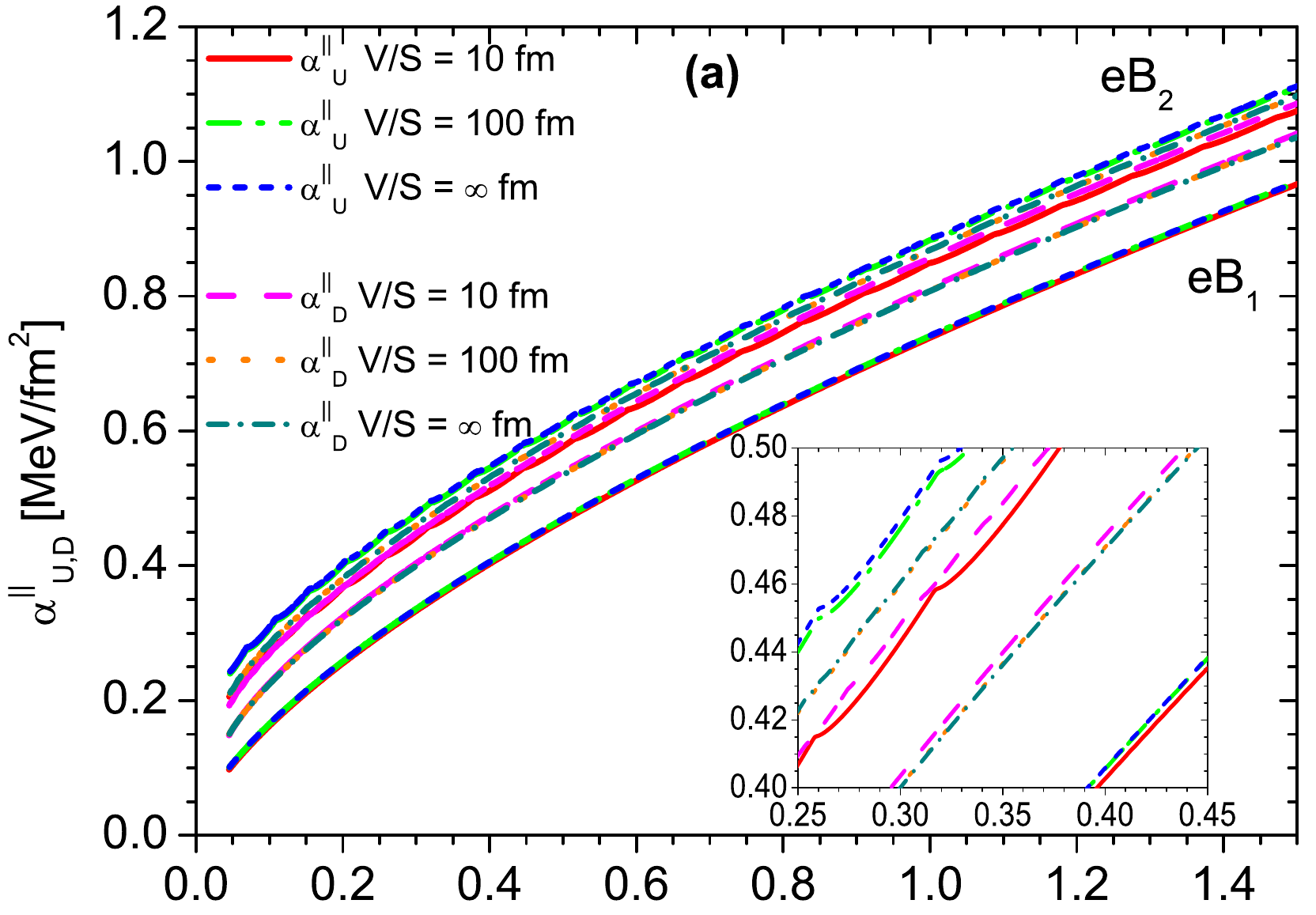}
\includegraphics[angle=0,scale=0.48]{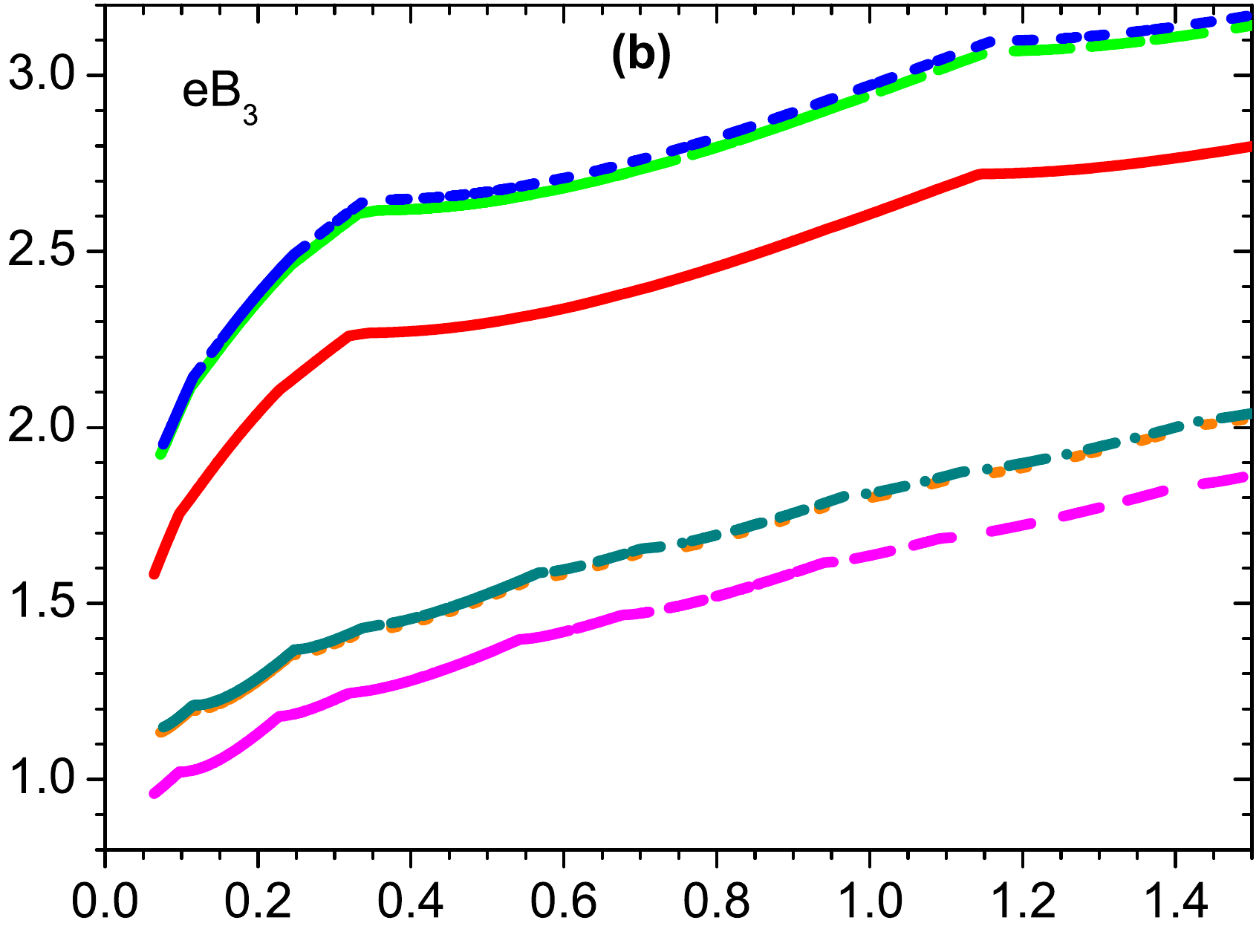}
\includegraphics[angle=0,scale=0.485]{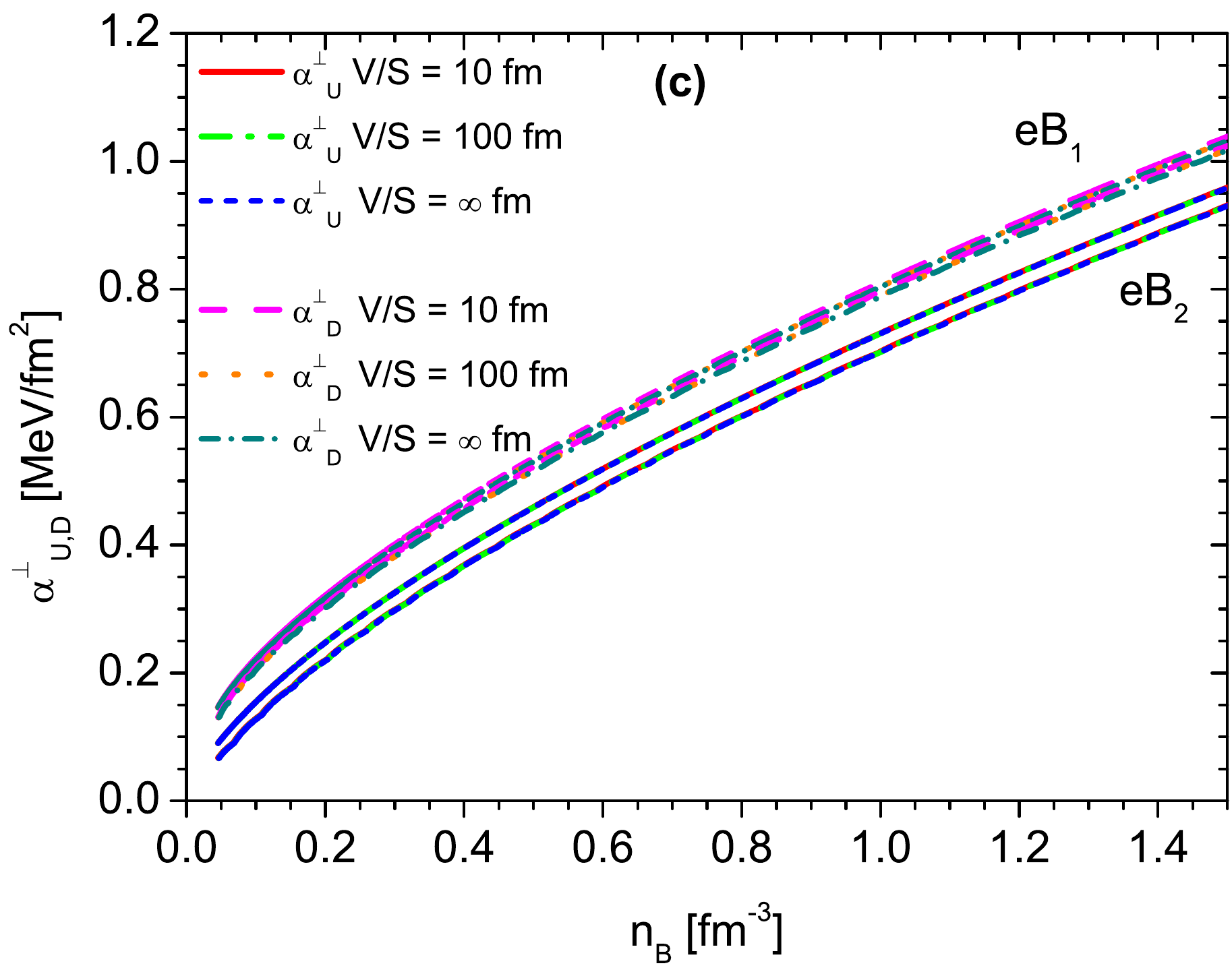}
\includegraphics[angle=0,scale=0.485]{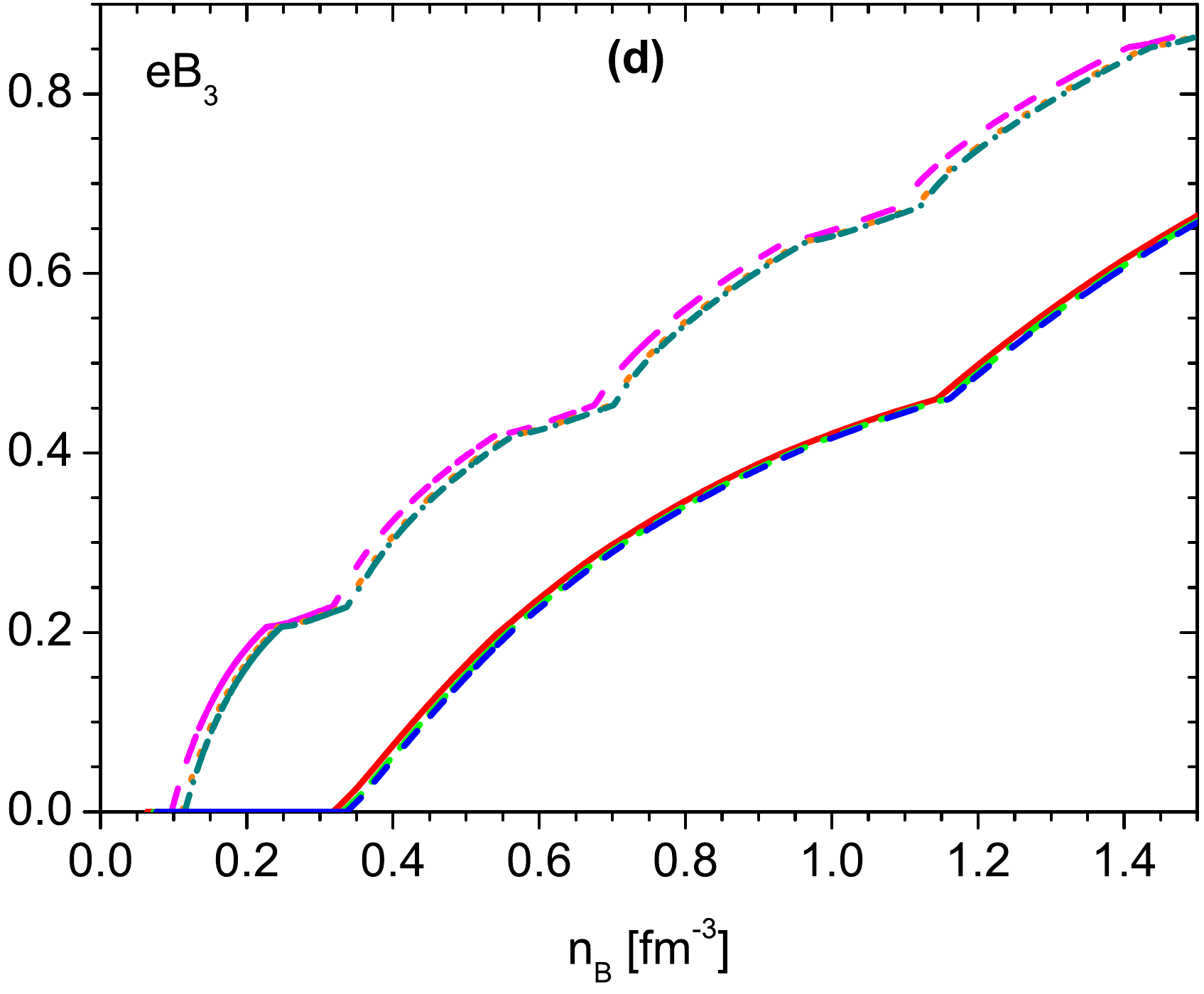}
\caption{Surface tension for quarks $u$ and $d$ in the parallel direction (panels (a) and (b)) and in the transverse direction  (panels (c) and (d)).   The values of $eB$ are the same as in the previous figure. }
\label{fig3}
\end{figure*}

\subsection{Transverse surface tension}

For obtaining the transverse surface tension contribution per flavor, the procedure is similar. In this case we start with the pressure of degenerate particles transverse to the field lines in bulk matter ${P}_f^{\perp}$, given in Eq. (15) of Ref.  \cite{Strick}, 
\begin{equation}
{P}_f^{\perp}
   = \frac{|q_f e B|^2}{2 \pi^2}
\sum_{\nu=0}^{\nu_{max}} \alpha_{\nu} \nu  \int_{0}^{k^{zF}_{f,\nu}}  \frac{dk_z}{\sqrt{k^2 +  m_f^2}}   ,
\end{equation}
and include the MRE density of states  $\rho_{MRE}$. Then,
\begin{equation}
- \Omega_f^{\perp}  =   \frac{V  |q_f e B|^2}{2 \pi^2}
\sum_{\nu=0}^{\nu_{max}} \alpha_{\nu} \nu \, \int_{\Lambda_{f,\nu}}^{k^{zF}_{f,\nu}}  \frac{\rho_{MRE} dk_z}{\sqrt{k^2 + m_f^2}}  
\end{equation}
Again, the latter expression has the form 
\begin{equation}
\Omega_f^{\perp}  = -  \Pi_f^{\perp}  V +  \alpha_f^{\perp}   S .
\end{equation}
Therefore, the transverse  pressure $\Pi_f^{\perp}$ and the transverse surface tension $\alpha_f^{\perp}$ within the MRE formalism  are given by:
\begin{eqnarray}
\Pi_f^{\perp}  &=&  \frac{|q_f e B|^2}{2 \pi^2} \sum_{\nu=0}^{\nu_{max}} \alpha_{\nu} \nu \, \int_{\Lambda_{f,\nu}}^{k^{zF}_{f,\nu}}  \frac{dk_z}{\sqrt{k^2 + m_f^2}}   ,  \\
\alpha_f^{\perp} & = & - |q_f e B|^2 \sum_{\nu=0}^{\nu_{max}} \alpha_{\nu} \nu  \int_{\Lambda_{f,\nu}}^{k^{zF}_{f,\nu}}  \frac { f_S(k) dk_z}{k  \sqrt{k^2 + m_f^2}}  . 
\end{eqnarray}

\subsection{Chemical equilibrium and charge neutrality}

In the present work we focus in the calculation of the surface tension of droplets of charge neutral quark matter in equilibrium under weak interactions. Chemical equilibrium is maintained by weak interactions among quarks, e.g.
$d \leftrightarrow u + e^- + \bar{\nu}_e$, $s \leftrightarrow u +
e^- + \bar{\nu}_e$, $u + d \leftrightarrow u + s$, from which we obtain the following relations between the chemical potentials:
\begin{eqnarray}
\mu_d &=& \mu_u + \mu_e, \\
\mu_s &=& \mu_d  .
\end{eqnarray}
Here we consider cold deleptonized matter; i.e. neutrinos leave freely the system. The charge neutrality condition reads
\begin{eqnarray}
\frac{2}{3} n_u - \frac{1}{3} n_d - \frac{1}{3} n_s - n_e = 0.
\label{charge_neutrality}
\end{eqnarray}
The number densities in the latter equation, can be obtained starting from \cite{Strick}
\begin{eqnarray}
n_f = \frac{|q_f e B|}{2\pi^2}  \sum_{\nu=0}^{\nu_{max}} \alpha_{\nu}   \int_{0}^{k^{zF}_{f,\nu}} dk_z  ,
\end{eqnarray}
and including the MRE density of states. Therefore, we obtain:
\begin{eqnarray}
n_f = \frac{|q_f e B|}{2\pi^2}  \sum_{\nu=0}^{\nu_{max}} \alpha_{\nu}  \int_{\Lambda_{f,\nu}}^{k^{zF}_{f,\nu}}  \left(  1 + \frac{2 \pi^2 S}{k V} f_S(k)     \right)  dk_z   .
\end{eqnarray}

\section{Results}

\begin{figure*}[htb]
\includegraphics[angle=0,scale=0.46]{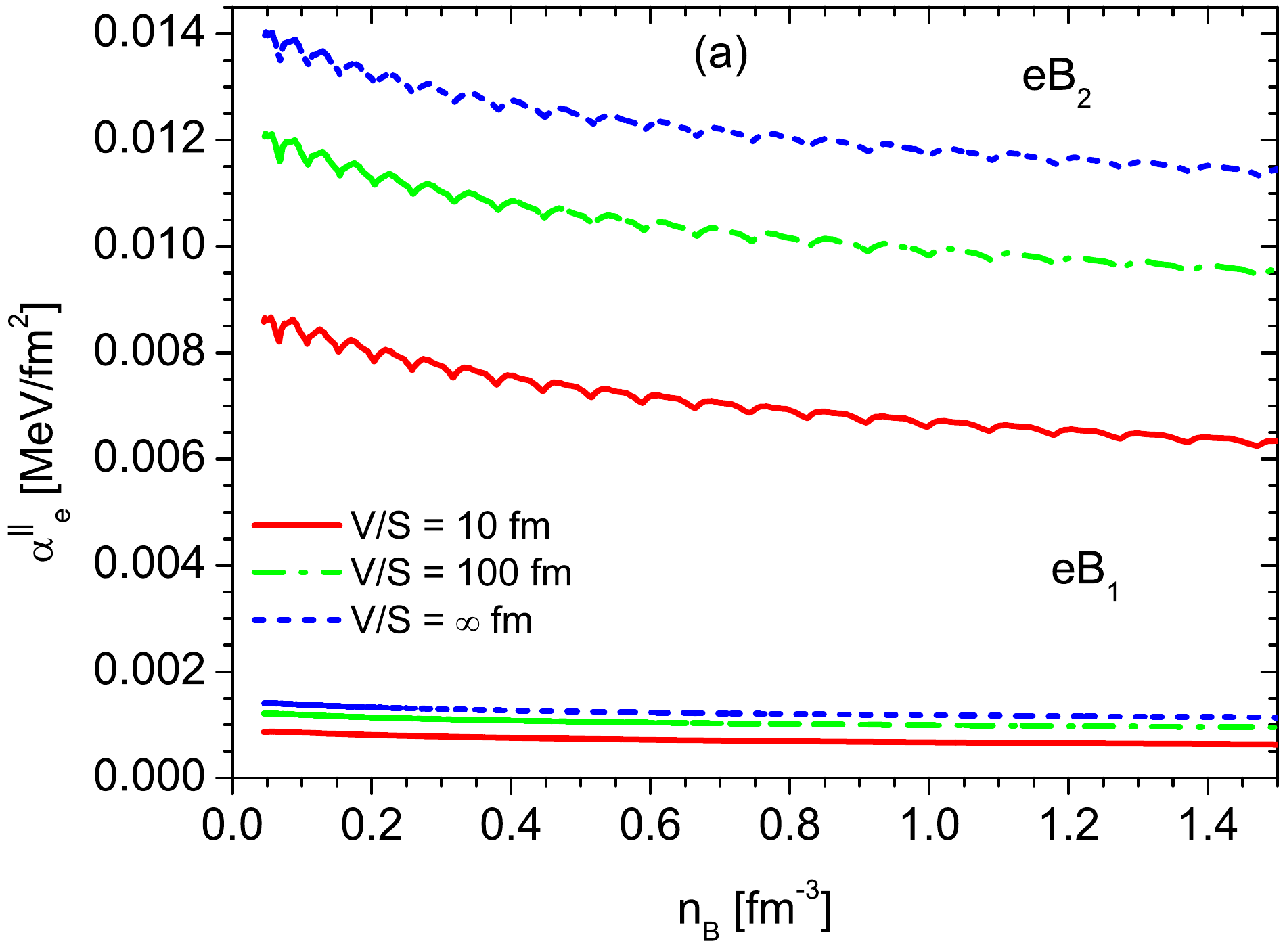}
\includegraphics[angle=0,scale=0.46]{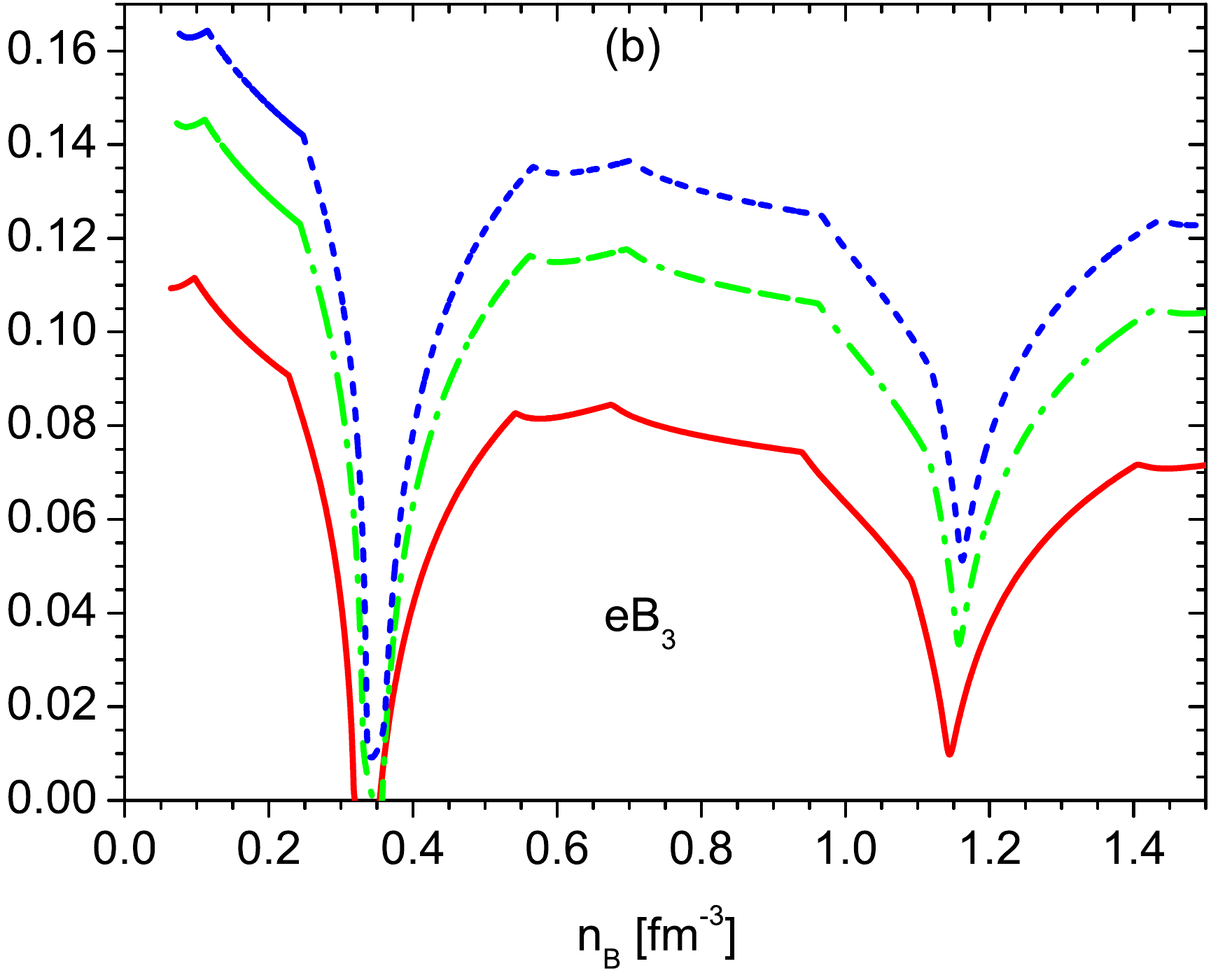}
\caption{Parallel surface tension for electrons. The transverse surface tension is not shown because it is negligible. }
\label{fig4}
\end{figure*}

\begin{figure*}[htb]
\includegraphics[angle=0,scale=0.46]{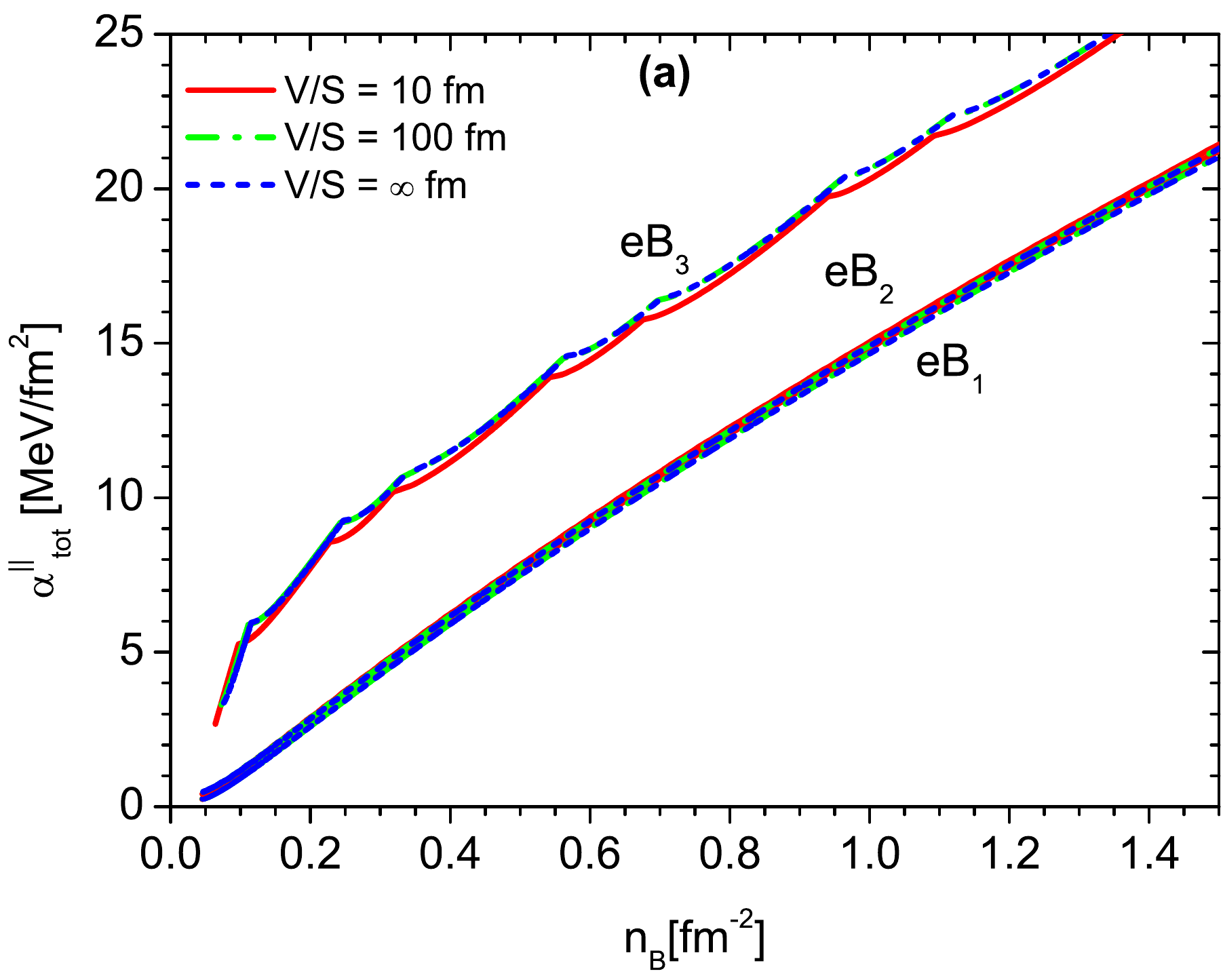}
\includegraphics[angle=0,scale=0.46]{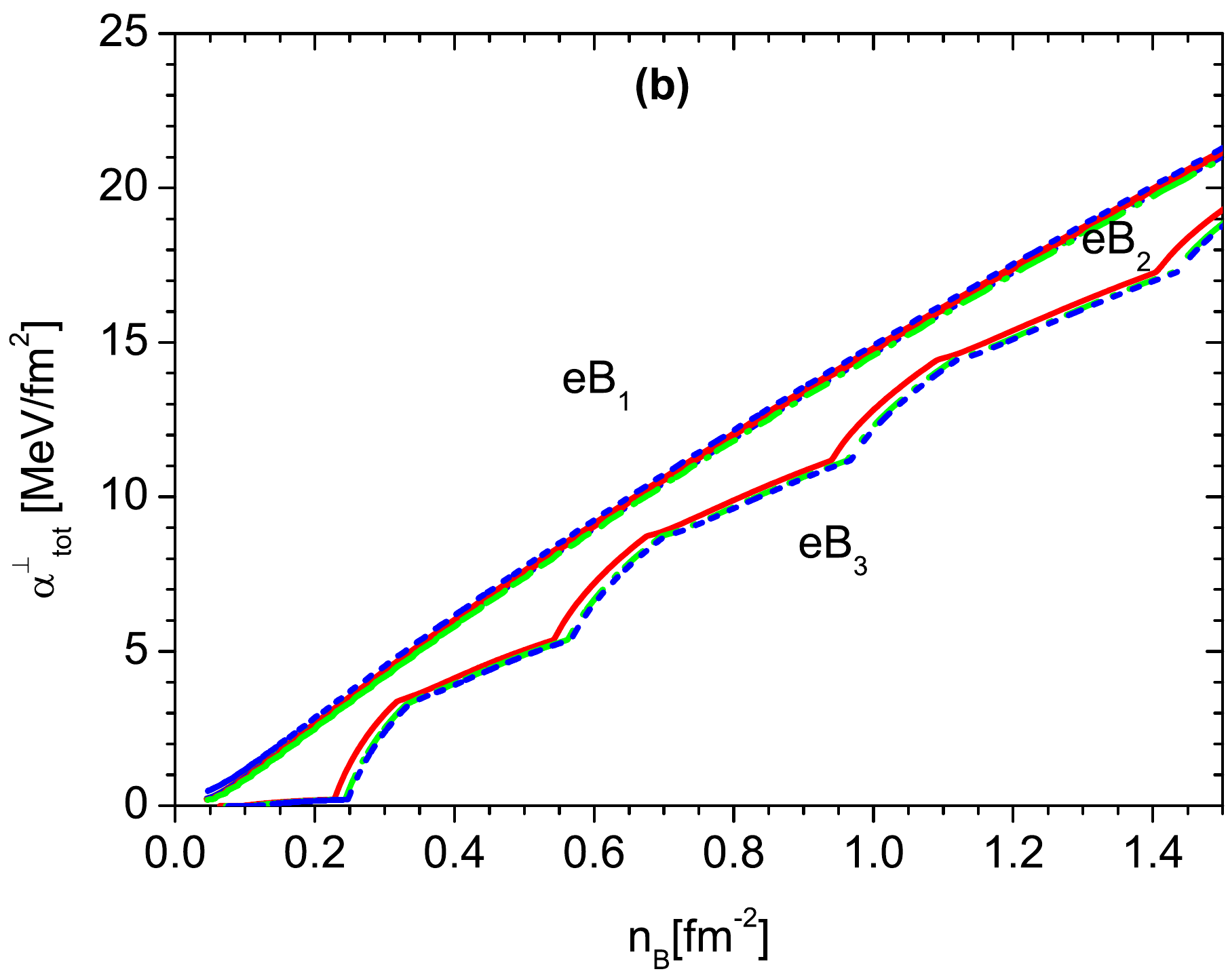}
\caption{Total surface tension obtained by summing the contributions of  quarks $u$, $d$, $s$ and electrons  in the parallel direction (panel (a)) and in the transverse direction  (panel (b)). }
\label{fig5}
\end{figure*}

In this section we calculate the surface tensions  $\alpha_f^{\parallel} $ and $\alpha_f^{\perp}$, as functions of the baryon number density $n_B = \frac{1}{3} (n_u + n_d + n_s)$ for different magnetic field intensities and different values of $V/S$. In the case of a spherical drop we would have $V/S = R/3$. However, in the presence of a strong magnetic field we expect the drop to be prolate, i.e. with a radius in the direction parallel to the magnetic field larger than the radius in the direction transverse to the magnetic field. The exact form of the drop as a function of the magnetic field will be explored in a future work. For the moment, we notice that $\alpha_f^{\parallel}$ and $\alpha_f^{\perp}$ don't depend on the exact geometry of the drop but only on the ratio $V/S$, which is taken here as a free parameter.

In Figs. \ref{fig2}$-$\ref{fig4} we show the contribution of quarks $u$, $d$, $s$ and electrons to the surface tension in the parallel and transverse directions for drops with $V/S$ =  10 fm, 100 fm and  for the bulk limit with $V/S = \infty$.  The largest contribution comes from the strange quark which may have a surface tension in the $\perp$ and $\parallel$ directions as large as $\sim 20 \, \mathrm{MeV/fm}^2$ for  baryon number densities and magnetic field intensities typical of neutron star's interiors (see Fig. \ref{fig2}). 
The contribution of quarks $u$ and $d$ is much smaller, with values that never exceed $\sim 3 \, \mathrm{MeV/fm}^2$ (see Fig. \ref{fig3}) and that of electrons is negligible in quark matter, with values below $\sim 0.15 \, \mathrm{MeV/fm}^2$ (see Fig. \ref{fig4}).

For fixed $eB$ and $V/S$, the surface tension is an increasing function of $n_B$. For ultra high values of the magnetic field (e.g. $eB = 5 \times 10^{-2} $ GeV$^2$) the curves clearly show de Hass$-$van Alphen oscillations related to the filling of new Landau levels. For lower fields, such oscillations are also apparent for the lightest particles (see e.g.   panel (a) of  Fig.  \ref{fig4}   and inset plot in panel (a) of Fig.  \ref{fig3}).

For fixed $n_B$ and $V/S$, the surface tensions in the $\perp$ and $\parallel$ directions are almost unaffected by the magnetic field if $eB$ is below $\sim  5 \times 10^{-3} $ GeV$^2$; as a consequence we have $\alpha_f^{\parallel} \approx \alpha_f^{\perp}$ below such $eB$.  For higher values of $eB$, e.g. $eB = 5 \times 10^{-2} $ GeV$^2$,  there is  a significant increase in  $\alpha_f^{\parallel}$ and a significant decrease in $\alpha_f^{\perp}$ with respect to the unmagnetized case, as can be seen in Figs. \ref{fig2}$-$\ref{fig4}.

Finally, for fixed $eB$ and $n_B$, the effect of varying $V/S$ is negligible for heavier particles, like strange quarks (see Fig. \ref{fig2}). However, for light quarks ($u$ and $d$) the parallel surface tension is significantly decreased with respect to the bulk case, but this happens only for $eB = 5 \times 10^{-2} $ GeV$^2$ (see panel (b) of Fig. \ref{fig3}). The transverse surface tension is almost unaffected by $V/S$  for any value of $B$. In the case of electrons, the effect of $V/S$ is even larger, because they are lighter. 

In Fig.  \ref{fig5} we show the total surface tension in the $\perp$ and $\parallel$ directions: 
\begin{eqnarray}
\alpha_{\mathrm{tot}}^{\parallel} & = & \sum_{i=u, d, s, e} \alpha_{i}^{\parallel} ,  \label{total1}\\
\alpha_{\mathrm{tot}}^{\perp} & = & \sum_{i=u, d, s, e} \alpha_{i}^{\perp}  .   \label{total2}
\end{eqnarray}

Since, in practice, the surface tension of quark matter in chemical equilibrium under weak interactions is largely dominated by strange quarks, the results in Fig.  \ref{fig5}  resemble those given Fig.  \ref{fig2}.

\section{Summary and conclusions}

In this work we have studied the surface tension of magnetized quark matter within the formalism of multiple reflection expansion (MRE). Quark matter is described as a mixture of free Fermi gases composed by quarks $u$, $d$, $s$ and electrons, in chemical equilibrium under weak interactions. Due to the presence of strong magnetic fields the transverse motion of these particles is quantized into Landau levels, and as a consequence, the surface tension has a different value in the parallel and transverse directions with respect to the magnetic field. 

We have calculated the surface tension in the $\perp$ and $\parallel$ directions for quark matter drops with different sizes.  Due to the magnetic field effect, such drops are expected to be prolate. However,  $\alpha_f^{\parallel}$ and $\alpha_f^{\perp}$  depend on the shape of the drop only through the ratio $V/S$. Such behavior arises because the  MRE density of states is smoothed to eliminate its fluctuating part  \cite{Balian1970}, and more specific details on the shape are washed out. In the present work, $V/S$ is taken as a free parameter with values 10 fm, 100 fm and  $\infty$.

The largest contribution to the surface tension comes from strange quarks which have a surface tension an order of magnitude larger than the one for $u$ or $d$ quarks and more than two orders of magnitude larger than for electrons. 
Our results show that the effect of varying $V/S$ is negligible for the strange quark. Therefore, although $V/S$ has a significant effect on the surface tension of light particles, the total surface tension given in Eqs. (\ref{total1})$-$(\ref{total2}) is insensitive to $V/S$ and depends mostly on the magnetic field and the baryon number density.  We also find that the surface tensions in the $\perp$ and $\parallel$ directions are almost unaffected by the magnetic field if $eB$ is below $\sim  5 \times 10^{-3} $ GeV$^2 \approx 0.25 m^2_{\pi}$, being $m_{\pi}$ the mass of the pion. Nevertheless, for higher values of $eB$,   there is  a significant increase in the parallel surface tension and a significant decrease in the transverse one with respect to the unmagnetized case (see in Fig. \ref{fig5}). 

In a recent work, Garcia and Pinto \cite{Garcia2013} considered magnetized two flavor quark matter within the Nambu-Jona-Lasinio model, and used a geometric approach for evaluating the surface tension. For unmagnetized matter, they find a surface tension  $\sim 30$ MeV fm$^{-2}$, which is significantly larger than the values found here for $u$ and $d$ quarks.  However, this is not surprising since calculations of the surface tension using the MRE formalism within the NJL model also give very large values \cite{Lugones2011,Lugones2013}; i.e. it is already known that the MIT bag model tends to give in general a significantly smaller surface tension than the NJL model. For magnetized matter, the results in Ref. \cite{Garcia2013} show that the surface tension oscillates slightly around the $B = 0$ value, for $0 < eB < 4m^2_{\pi}$. However, for $4m^2_{\pi} < eB <6m^2_{\pi}$ it decreases  reaching a minimum value at $eB \approx 6 m^2_{\pi}$ which is about $30 \%$ smaller than the $B = 0$ result. For larger magnetic fields, the surface tension increases continuously, reaching a value of $\sim 40$ MeV fm$^{-2}$ for $eB \approx 10 m^2_{\pi}$. There are some significant differences between the results  of Ref. \cite{Garcia2013} and ours.  First, we derive two surface tensions, one parallel and another transverse to the magnetic field, and they have a qualitative different dependence on $B$:  the parallel surface tension increases with $B$ and the transverse one decreases.  Secondly,  our results are more sensitive to an increase of the magnetic field. For example, for $eB_3 = 5 \times 10^{-2} $ GeV$^2$ $\approx 2.5 m^2_{\pi}$, Ref. \cite{Garcia2013}  finds  a  variation of a few percent in the surface tension with respect to the $B=0$ case.  In our calculations, the difference with respect to the unmagnetized case is in the range $\sim 20-200 \%$ for typical compact star densities.

Another issue that needs some discussion is the effect of the Coulomb interactions. Notice that we are enforcing local charge neutrality in Eq. (\ref{charge_neutrality}). Such approximation is valid for drops that are large compared with the Debye screening length $\lambda_D$.  However, for drops whose size is of the order or smaller than $\lambda_D$, electric fields may have a significant effect and global charge neutrality should be taken into account properly. Since the Debye screening length in quark matter is typically of the order of 5 fm \cite{Heiselberg1993,Alford2006}, we expect that our results are reliable for the cases with $V/S = 100, \infty$ fm.  However, the case with $V/S = 10$ fm (which would correspond to a radius $R=3V/S=30$ fm if the drop were spherical)  requires a more sophisticated analysis including the Coulomb potential and solving consistently the Poisson equation together with the other equations.  Such analysis is beyond the scope of the present work, but, in spite of the problem being highly non-linear, some approximate conclusions can be foreseen in the light of previous works \cite{Endo2006,Alford2006}. In Ref. \cite{Endo2006}, the Coulomb interaction effect is consistently taken into account in the hadron-quark mixed phase, which consists of an equilibrium configuration of quark drops embedded in a hadronic medium. It is found that there is a rearrangement of the charged particles near the drop boundary: the negatively charged particles in the quark phase (quarks $d$, $s$ and electrons) are attracted toward the boundary and the positively charged particles (quarks $u$) are repelled from the boundary  \cite{Endo2006,Alford2006}. Such effect may affect the surface tension indirectly because it induces a change in the particle abundances.  Since the surface tension is largely dominated by the contribution of the strange quark, an increase in the $s$ quark number density near the boundary could lead to a surface tension that is somewhat larger that the here-found values.  However, such effect is expected to be small, because the net charge in the quark drop is shown to be small due to screening effects \cite{Endo2006}. 

Our  results may have interesting consequences for neutron star interiors. For example, according to \cite{Voskresensky2003}, beyond a limiting value of $\alpha \approx 65$ MeV/fm$^2$ the structure of the mixed quark-hadron phase that may form inside a neutron star becomes mechanically unstable  and local charge neutrality is recovered, leading to a sharp interface.  Our results show that although the magnetic field can change significantly the  surface tension of quark matter, its effect wouldn't be large enough as to change the nature of the interface from mixed to sharp. However, since quark matter drops adopt a prolate shape inside a strong magnetic field, we may expect significant changes in the geometrical structure of the drops, rods, slabs and all the ``pasta'' phase configurations that have been conjectured to exist if global charge neutrality is allowed.   

Another relevant consequence is related to the triggering of the conversion of a hadronic star into a quark star (hybrid or strange). The conversion is expected to begin with the nucleation of tiny quark matter drops at the core of the hadronic star which thereafter may grow to a macroscopic size and initiate a combustion front \cite{Lugones2016}. Since strong magnetic fields modify the surface tension, we may expect changes in the critical spectrum of fluctuations \cite{Lugones2011}, in the nucleation rate, and probably, asymmetric combustion fronts. These effects will be explored in future work.

\end{document}